\documentclass[aps,pra,superscriptaddress,reprint,floatfix,longbibliography,showpacs,amsfonts,amsmath,amssymb]{revtex4-1}
\DeclareUnicodeCharacter{2009}{\,}
\usepackage{graphicx}
\usepackage{hyperref}
\usepackage{xr-hyper}
\usepackage{upgreek}
\usepackage{epstopdf}
\usepackage[note-name=, use-sort-key = false]{notes2bib}
\usepackage{color}
\usepackage{soul}
\usepackage{ulem}

    \hypersetup{
    	colorlinks = true,
		citecolor = blue,
		bookmarksnumbered = true,
		urlcolor = blue
	}

\renewcommand{\thefigure}{\arabic{figure}}

\begin{document}

\title{Universality of magnetic susceptibility in the conical state of kagome ferromagnet Fe$_3$Sn$_2$}

\author{Lilian Prodan}
\email{lilian.prodan@uni-a.de}
\affiliation{Experimentalphysik V, Center for Electronic Correlations and Magnetism, Institute of Physics, University of Augsburg, D-86159 Augsburg, Germany}

\author{ Donald M. Evans}
\affiliation{Experimentalphysik V, Center for Electronic Correlations and Magnetism, Institute of Physics, University of Augsburg, D-86159 Augsburg, Germany}
\affiliation{Department of Sustainable Energy Technology, SINTEF Industry, Oslo, Norway}

\author{ Lukas Puntigam}
\affiliation{Experimentalphysik V, Center for Electronic Correlations and Magnetism, Institute of Physics, University of Augsburg, D-86159 Augsburg, Germany}

\author{Istv\'an K\'ezsm\'arki}
\affiliation{Experimentalphysik V, Center for Electronic Correlations and Magnetism, Institute of Physics, University of Augsburg, D-86159 Augsburg, Germany}

\author{Vladimir Tsurkan}
\affiliation{Experimentalphysik V, Center for Electronic Correlations and Magnetism, Institute of Physics, University of Augsburg, D-86159 Augsburg, Germany}
\affiliation {Institute of Applied Physics, Moldova State University, MD 2028, Chișinău, R. Moldova}

\date{\today}

\begin{abstract}
We report universal behavior of the differential magnetic susceptibility (DMS) in the conical phase that mediates the spin-reorientation (SR) transition of the kagome ferromagnet Fe$_3$Sn$_2$. Within the SR temperature range, the DMS isotherms exhibit extremely narrow crossing regions, forming isosbestic points. Using an isosbestic-invariance analysis, we show that the isotherms collapse onto a single temperature-independent curve, revealing quadratic-in-temperature corrections to the susceptibility. Complementary field-dependent magnetic-force-microscopy measurements uncover evolution of spin textures from stripe-like domains at low fields to isolated bubble-like domains near the isosbestic field ($\sim 0.6$~T), a behavior not previously reported in bulk Fe$_3$Sn$_2$ within the conical state. These findings point to a universal mechanism for the emergence of complex magnetic textures near isosbestic points, driven by the competition between magnetocrystalline anisotropy, dipolar interactions, and external magnetic field.
\end{abstract}

\maketitle
\section{Introduction}
The isosbestic invariance approach provides a general framework for analyzing a measurable physical quantity $Q(x_1,x_2,\ldots,x_n)$ that depends on multiple variables~\cite{Vollhardt1997,Greger2013}. By parameterizing one variable, one can uncover the functional dependence of $Q$. In certain cases, the resulting curves intersect within a narrow region or even at a singular point, referred to as an isosbestic point~\cite{Vollhardt1997}. Such crossing points have been identified in a wide range of material properties, including specific heat, optical conductivity, dielectric response, Raman spectra, and magnetoresistance~\cite{Biswas2020, Lunkenheimer2017, Ibarra2021, Volkov2021, Orendac2017, Grafe2017, Anisimov2025, Wang2014, Reschke2017}. They often reflect universal constraints such as entropy conservation, spectral-weight redistribution, or sum rules~\cite{Vollhardt1997,Greger2013}. The isosbestic approach is therefore a powerful tool for extracting leading parameter dependence and uncovering hidden feature of complex physical systems.

Here, we apply this concept to the magnetic susceptibility across the spin-reorientation transition in the kagome ferromagnet Fe$_3$Sn$_2$. This compound belongs to the broad family of kagome topological magnets, which have attracted strong interest because of the interplay between geometric frustration, spin--orbit coupling, and electronic correlations~\cite{Fenner2009,Yin2018,Ye2019}. These interactions generate unconventional electronic structures with Dirac bands, flat bands, and large Berry curvature, strongly affecting magnetic, charge, and thermal transport properties~\cite{Wang2016,Ye2018,Yin2019,Biswas2020,Lin2018a,Schilberth2022,Du2022,Du2020}. Fe$_3$Sn$_2$ has emerged as a prototypical itinerant kagome ferromagnet in which topological electronic states coexist with robust magnetic order. It is also known to host topologically protected skyrmion bubbles at room temperature, whose stability and morphology are highly sensitive to geometrical confinement, strain, and light~\cite{Hou2018,Hou2018a,Altthaler2021,Kong2023,Kovacs2025}. This remarkable tunability reflects the delicate balance among magnetocrystalline anisotropy, dipolar interactions, and magnetoelastic coupling, making Fe$_3$Sn$_2$ a model system for studying controllable spin textures and field-driven magnetic states.

The temperature-driven spin reorientation in Fe$_3$Sn$_2$ has been known for decades~\cite{Caer1978,Caer1979,Malaman1978}. Early neutron diffraction and M\"ossbauer studies established that the Fe moments align along the $c$ axis at 300~K, whereas below 60~K they lie predominantly within the kagome $ab$ plane. Subsequent studies using magnetization, $ac$ susceptibility, magnetic force microscopy (MFM), optical and Raman spectroscopies, and magnetotransport measurements provided extensive phenomenology of this transition~\cite{Biswas2020,Fenner2009,Heritage2020,He2021,Wu2021,Kumar2019}. Our recent studies on single crystals of Fe$_3$Sn$_2$ clarified the nature of this transition by identifying an easy-cone state that mediates the spin reorientation (SR)~\cite{Prodan2025}. In this intermediate regime, the order parameter, represented by the reorientation angle $\theta$, evolves continuously between 130~K and 40~K and is governed by the competition between first- and second-order magnetocrystalline anisotropy constants. 
However, despite these efforts, the field evolution of magnetic response across the SR regime remains poorly understood.

In this work, we address these questions through combined bulk magnetization and microscopic imaging measurements. We show that the differential magnetic susceptibility isotherms exhibit well-defined crossing points and obey a universal relation within the spin-reorientation regime. Complementary MFM measurements further reveal that bubble-like magnetic domains emerge in fields close to the isosbestic point, establishing a direct connection between bulk magnetic response and field-driven real-space spin-texture transformations.

\begin{figure} [t!]
 \centering
     \includegraphics[scale=0.092]{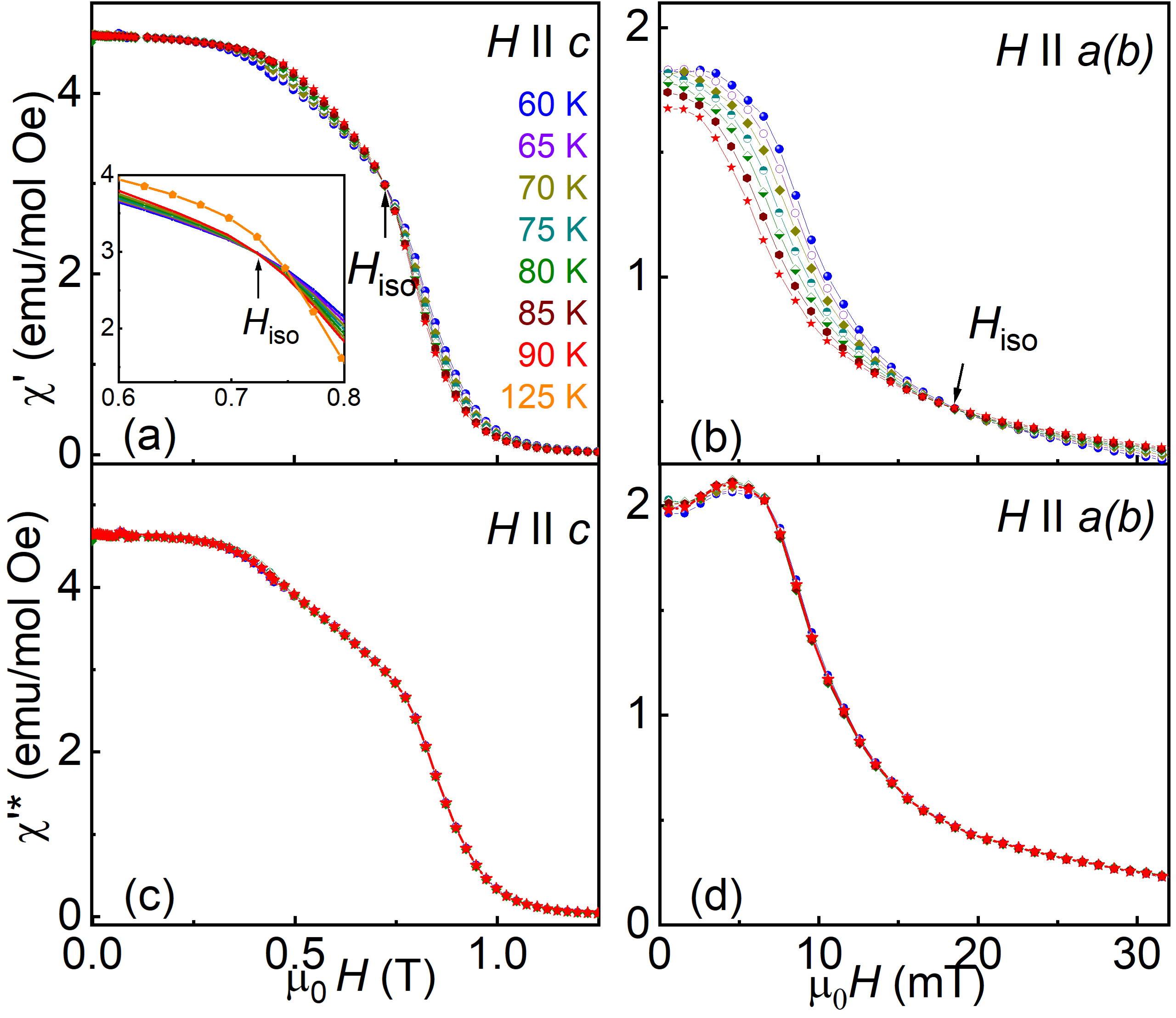}
     \caption{Differential magnetic susceptibility $\chi'$ of single-crystalline Fe$_3$Sn$_2$ (sample S1) as a function of magnetic field $H$ applied along the $c$ axis (a) and the $a$ axis (b). Panels (c) and (d) show the scaled quantity $\chi_i^{*}$, demonstrating the collapse of the corresponding isotherms onto a single temperature-independent curve. The color code in panel (a) applies to all panels. The inset to panel (a) shows an enlarged view highlighting the deviation of the 125~K isotherm from those measured between 60 and 90~K. $H_{\mathrm{iso}}$ denotes the isosbestic field. }
     \label{fig:Fig1}
 \end{figure}

\section{Results and discussion}
To investigate this behavior experimentally, we used single crystals of Fe$_3$Sn$_2$ grown by the chemical transport reaction method~\cite{Prodan2025}. Figures~1(a) and 1(b) show the field-dependent differential magnetic susceptibility (DMS) isotherms, $\chi'(H)=dM/dH$, for one representative single-crystalline sample (S1), measured between 60 and 90~K. The data are presented for two orientations of the magnetic field $H$, applied along the $c$ and $a$ axes. The striking feature of the $\chi'(H)$ dependence is a very narrow field region approaching a singular point, which we denote as the isosbestic field $H_{\mathrm{iso}}$, where all isotherms intersect. With increasing temperature, for $H \parallel c$, $\chi'(H)$ exhibits complementary behavior below and above $H_{\mathrm{iso}}$: $\chi'$ decreases for $H < H_{\mathrm{iso}}$ and increases for $H > H_{\mathrm{iso}}$. For $H \parallel a$, the field dependence of $\chi'$ shows the opposite trend. This contrasting behavior for the in-plane and out-of-plane field configurations indicates a redistribution of the magnetization between the $c$ and $a$ directions, as expected across the spin-reorientation transition.

To quantify this universal crossing behavior, we analyzed the evolution of $\chi'$ with temperature and field using the isosbestic-invariance approach developed in Ref.2. The experimental data, i.e., the field dependence of $\chi'$ at different temperatures, were described using the following ansatz:
\begin{equation}
\chi'(H,T)=\chi'(H,0)+T^{2}\chi_2'(H),
\label{eq1}
\end{equation}
where the magnetic field $H$ is the variable and the temperature $T$ is the parameter. Note that Eq.~(\ref{eq1}) does not contain a term linear in temperature, since $\chi'$ is temperature independent at the isosbestic field ($d\chi'/dT \approx 0$).

The scaling function $\chi_2'(H)$ in Eq.~(\ref{eq1}) is defined as
\begin{equation}
\chi_2'(H)=\frac{\chi'(H,T_1)-\chi'(H,T_2)}{T_1^2-T_2^2}.
\label{eq2}
\end{equation}

The validity of Eq.~(\ref{eq1}) was checked through the temperature independence of the quantity
\begin{equation}
\chi_i^{*}(H,T)=\chi'(H,T)-T^2\chi_2'(H).
\label{eq3}
\end{equation}

In Eq.~(\ref{eq2}), we set $T_1=60$~K and $T_2=90$~K. These limits were chosen because outside this temperature range $\chi'(H,T)$ begins to deviate from the behavior near the isosbestic point, as shown in the inset of Fig.~1(a). The results of the scaling analysis are presented in Figs.~1(c) and 1(d). For both field orientations, the experimental curves collapse onto a single universal curve, $\chi_i^{*}(H,T)$. The collapse of the curves demonstrates that the leading temperature dependence of the susceptibility is captured by the quadratic term throughout the conical-state regime. 

The universality of $\chi_i^{*}(H)$ was further verified for other Fe$_3$Sn$_2$ samples. Figure~A1 shows the DMS and the corresponding scaled isotherms for sample S2, which exhibits magnetic behavior similar to that of S1. The only significant difference among the studied samples lies in the value of the isosbestic field $H_{\mathrm{iso}}$, which is mainly related to the demagnetization effect, as discussed below. 

To clarify the microscopic origin of the universal behavior observed in the DMS and to identify the physical mechanisms responsible for the emergence of the isosbestic point, it is essential to address several key questions concerning the field and temperature evolution of magnetic susceptibility. In particular, two of these fundamental questions were already formulated in Ref.~\cite{Vollhardt1997} and provide a natural framework for the present analysis:

(i) \textit{Why do the $\chi'(H)$ curves cross at all?}

(ii) \textit{Why is the crossing feature confined to a narrow region despite significant changes in temperature?}

The answer to the first question follows directly from the magnetic anisotropy of Fe$_3$Sn$_2$: the DMS isotherms cross due to the presence of sizable magnetocrystalline anisotropy in Fe$_3$Sn$_2$~\cite{Prodan2025,Fayyazi2019}. Note that for an isotropic ferromagnet, the DMS isotherms intersect only at two points: first at $H=0$, where the value is determined by the demagnetizing factor, and second in the saturated state. In the isotropic case, the DMS isotherms never cross at intermediate fields (see Appendix, Fig.~A2). In contrast, in anisotropic ferromagnets, the DMS curves cross due to the temperature dependence of the magnetocrystalline anisotropy. Figure~A3 illustrates this behavior for the anisotropic kagome ferromagnet Fe$_3$Sn, which possesses a kagome structure similar to that of Fe$_3$Sn$_2$~\cite{Prodan2023}.

The second question is also related to magnetocrystalline anisotropy, in particular to the temperature-induced sign change of the anisotropy constants $K$. As seen in Fig.~A3, despite the crossing of the DMS curves, an isosbestic point does not appear in Fe$_3$Sn because the temperature dependence of the anisotropy is dominated by the monotonic decrease of the first-order anisotropy constant $K_1$ with decreasing temperature, while the second-order anisotropy constant $K_2$ is significantly smaller than $K_1$ (see Ref.~\cite{Prodan2023}). As a result, the anisotropy energy does not change sign with temperature. 

\begin{figure} [t!]
 \centering
     \includegraphics[scale=0.1]{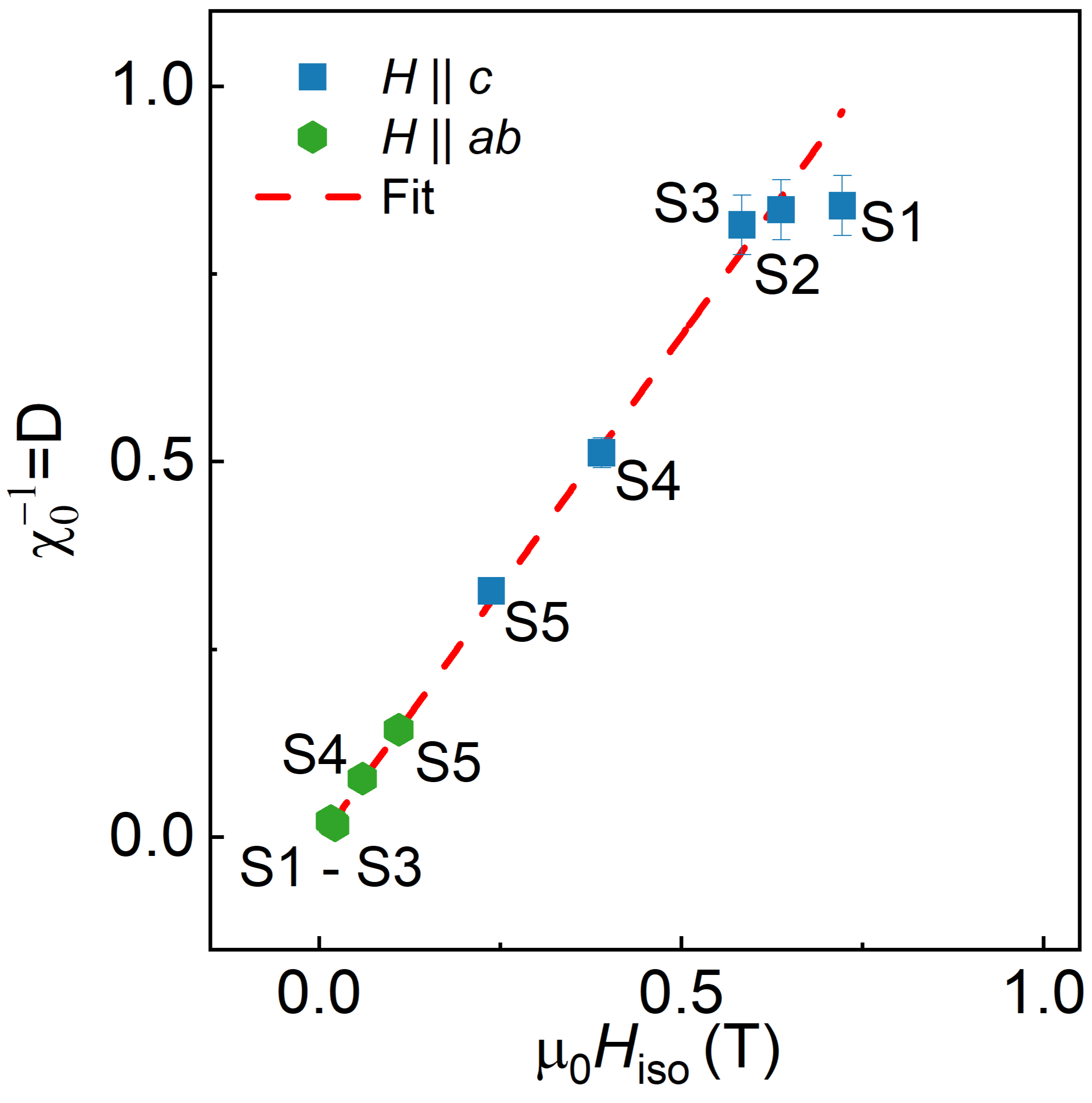}
     \caption{ Relation between the demagnetizing coefficient $D$ and the isosbestic field $H_{\mathrm{iso}}$ for the five studied samples (S1--S5). Large values of $D$ correspond to out-of-plane field configurations ($H~\parallel~c$), whereas small values correspond to in-plane configurations ($H~\parallel~ab$). The dashed line represents a linear fit to the data. }
     \label{fig:Fig2}
 \end{figure}

In the title compound Fe$_3$Sn$_2$, the spin-reorientation angle $\theta$ relative to the $c$ axis is governed by the balance between the first- and second-order magnetocrystalline anisotropy constants $K_1$ and $K_2$, which are of the same order of magnitude~\cite{Prodan2025}. Below 60~K, the system is in the easy-plane state, characterized by the ratio $|K_1/K_2|>2$ and $\theta=90^\circ$. At $T \approx 60$~K, this ratio decreases below 2 and the system enters the conical state. In this regime, $K_1$ is negative and its magnitude decreases with increasing temperature, while $K_2$ remains positive. The reorientation angle then follows the relation 
$\theta=\arcsin\left(\sqrt{\frac{-K_1}{2K_2}}\right)$ (see Ref.~\cite{Horner1968}). At 90~K, the ratio $K_1/K_2$ is close to $-1$, yielding a reorientation angle $\theta \sim 45^\circ$. At $T \sim 130$~K, Fe$_3$Sn$_2$ enters the easy-axis state as $K_1$ changes sign ($K_1=0$, $\theta=0^\circ$). Thus, the observed crossing point, between 60~-~90~K,  is located within the conical-state.

These considerations lead to the next question:

(iii) \textit{What is the physical meaning of the isosbestic field $H_{\mathrm{iso}}$, and what factors ultimately determine its value?}

To address this question, we analyzed the experimental data for samples with different shapes and found a clear correlation between the isosbestic field $H_{\mathrm{iso}}$ and the demagnetizing coefficient $D$~\footnote{The demagnetization coefficient $D$ was calculated from the initial slope of the magnetization curves, $\chi'(0)$, measured at 400~K for the out-of-plane configuration and at 2~K for the in-plane configuration, where the contribution of the magnetocrystalline anisotropy is negligible.}. Remarkably, $H_{\mathrm{iso}}$ exhibits an almost linear dependence on $D$ across samples for which the in-plane and out-of-plane demagnetizing factors differ by more than a factor of 50 (see Fig.~2). This striking proportionality indicates that $H_{\mathrm{iso}}$ contains a substantial contribution from the demagnetizing field $H_D = D M_{\mathrm{iso}}$, where $M_{\mathrm{iso}}$ is the magnetization at the isosbestic point. In a uniaxial crystal, the magnetization along the hard direction is governed by the combined action of the demagnetizing and anisotropy fields. For a magnetic field applied along the $c$ axis, saturation occurs at fields corresponding approximately to the sum of the demagnetizing field and the anisotropy field~\cite{Kalvius1979,Asti1974}. Assuming that similar relations hold at the isosbestic point, the constant value of the isosbestic field implies that the temperature-dependent variations of the demagnetizing field are compensated by corresponding changes in the anisotropy field associated with the spin reorientation, thereby stabilizing the universal crossing condition.

To obtain real-space evidence for the universality of the magnetic susceptibility, we imaged the evolution of the magnetic domain structure across the spin-reorientation transition. MFM measurements were performed on the as-grown $ab$ plane of sample S2 at various temperatures and magnetic fields applied along the $c$ axis (see Ref.~\cite{Prodan2025} for experimental details). Previous zero-field studies revealed a dendritic pattern of alternating domains pointing upward and downward along the $c$ axis~\cite{Heritage2020,Altthaler2021,Prodan2025,Fayyazi2019,Xie2024}. Below 155~K, a gradual transformation occurs from the dendritic state to predominantly stripe domains~\cite{Heritage2020,Prodan2025,Xie2024}. The stripe domains persist down to 70~K, indicating the presence of a sizable out-of-plane component of the magnetization. Only at 60~K the domain structure change significantly due to the almost in-plane alignment of the magnetization.

Figure~3 illustrates the evolution of the magnetic domain structure at 80~K with magnetic field applied along the $c$ axis. At 0.2~T, the branching of the stripe domains is strongly reduced. Upon increasing the field to 0.5~T, the stripe domains transform into a cocoon-like domain structure~\cite{Grelier2022} with a preferred in-plane orientation. At this field, isolated bubble-like domains begin to emerge. At 0.6~T, the domain pattern consists almost exclusively of isolated bubbles [Fig.~3(d) and 3(e)]. Notably, these bubble-like domains appear at fields close to the isosbestic field. At higher fields (0.8~T), the magnetic contrast largely vanishes, and the image mainly reflects the underlying surface morphology.

Previous room-temperature studies have shown that stripe domains in Fe$_3$Sn$_2$ can host topologically nontrivial spin textures, such as Bloch points and hybrid skyrmion bubbles (biskyrmions)~\cite{Altthaler2021,Hou2018,Kong2023}, which persist down to about 150~K in thin lamellae~\cite{Chen2021}. We therefore suggest that the bubble-like domains observed here at much lower temperatures may have a similar topological origin. Since these bubbles emerge from the gradual transformation of the stripe domains near the isosbestic field, they likely correspond to skyrmion-like textures stabilized by the balance among magnetocrystalline anisotropy, dipolar interactions, and external magnetic field. This observation suggests that the universal behavior of the differential magnetic susceptibility may be closely linked to the field-induced evolution of such topologically nontrivial spin textures.
 
The evolution of the magnetic structure across the SR transition in an external magnetic field close to $H_{\mathrm{iso}}$ is shown in Fig.~4. We observe a preferential in-plane orientation of the domains throughout the SR transition, similar to the behavior reported for nanometer-thick Fe$_3$Sn$_2$ plates at 300~K, where such textures are stabilized by tilting the sample with respect to the $c$ axis~\cite{Chen2021,Tang2020}. While a full understanding of the star-like patterns goes beyond the scope of this work, we note that the threefold rotational symmetry, particularly visible in Figs.~4(e) and 4(f), could reflect the underlying lattice symmetry combined with a slight misalignment of the crystal relative to the applied field. Together, these results demonstrate that spin reorientation creates natural conditions for the preferential in-plane orientation of magnetic domains through variations of the magnetocrystalline anisotropy and that such textures can be realized not only in lamellae but also in bulk samples.

\begin{figure} [t!]
 \centering
     \includegraphics[scale=0.09]{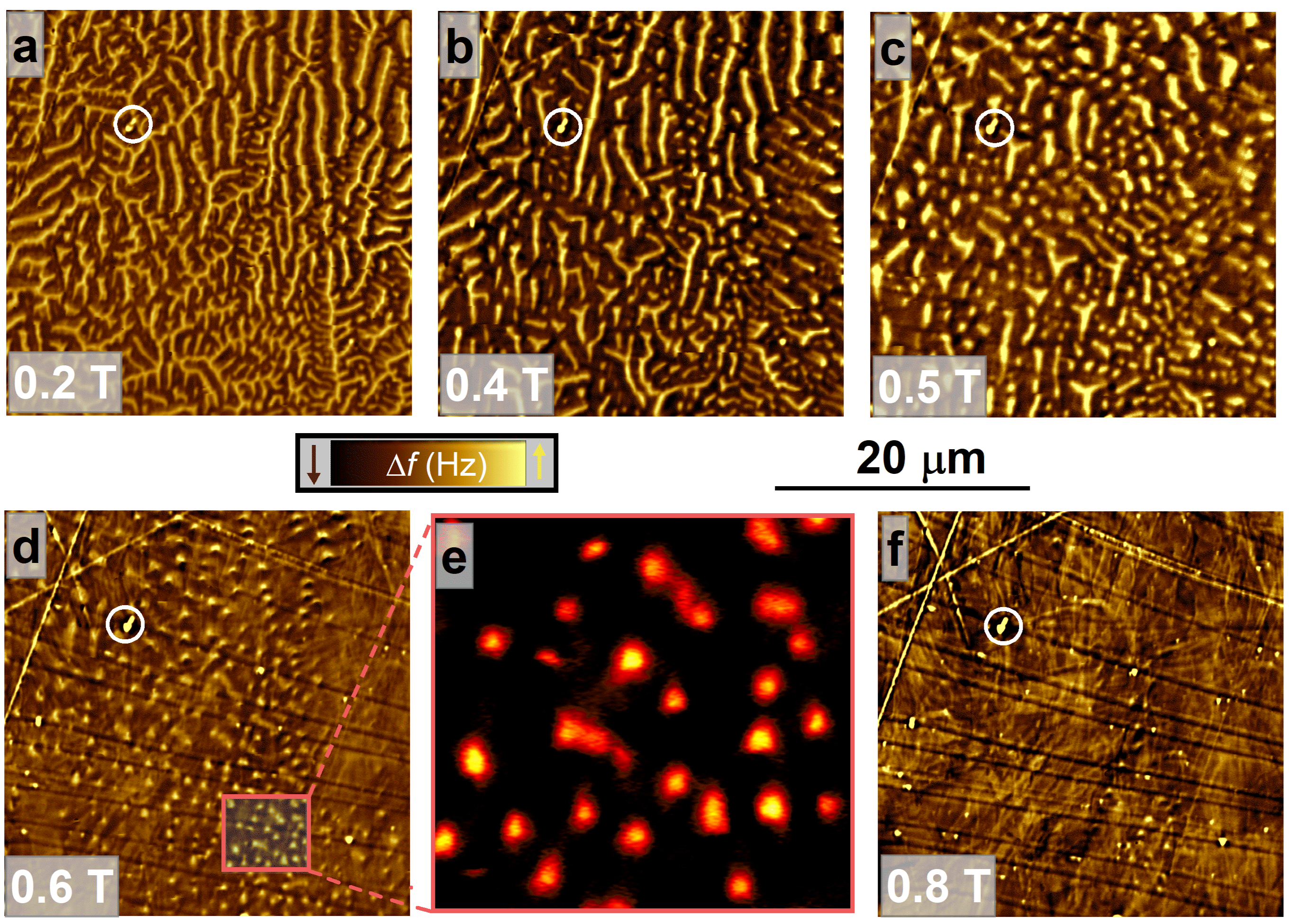}
     \caption{ Magnetic-field evolution of the domain patterns measured at 80~K. The inset between the panels indicates the upward (yellow) and downward (brown) magnetization directions. White circles mark a reference point defined by the surface morphology. Each scan covers an area of $30 \times 30~\mu\mathrm{m}^2$. Insets at the bottom left indicate the applied magnetic field. Panel (e) was obtained by subtracting the image in (f) from that in (d), revealing isolated bubble-like domains.}
     \label{fig:Fig3}
 \end{figure}

\begin{figure} [t!]
 \centering
     \includegraphics[scale=0.09]{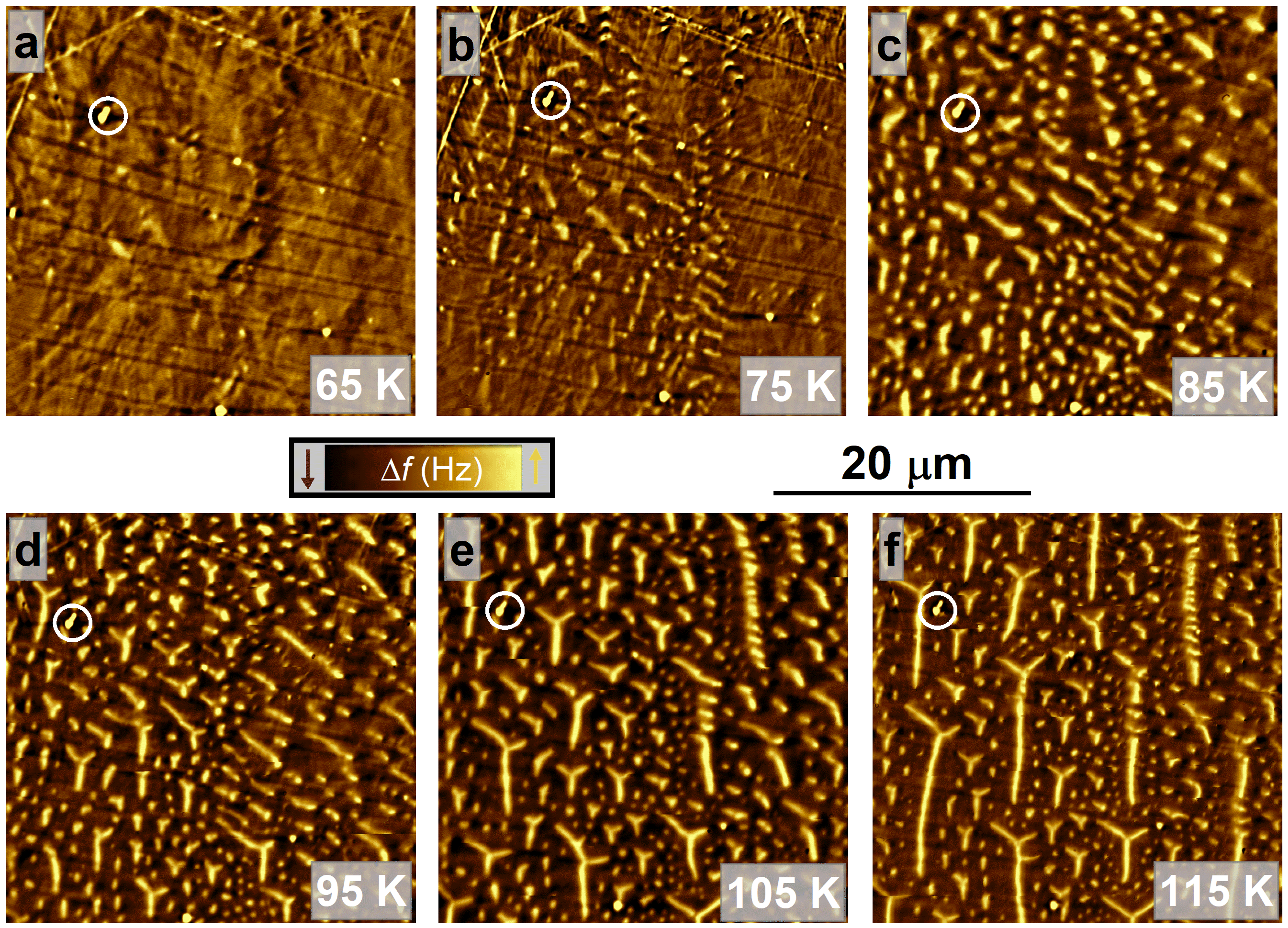}
     \caption{ Temperature-driven evolution of the domain patterns across the spin-reorientation transition in an applied magnetic field of 0.55~T. White circles mark a reference point defined by the surface morphology. Each image covers an area of $30 \times 30~\mu\mathrm{m}^2$. Insets at the bottom right indicate the measurement temperature. }
     \label{fig:Fig4}
 \end{figure}
 
\section{CONCLUSION}
In summary, our combined magnetization and MFM studies on bulk single crystals of Fe$_3$Sn$_2$ reveal distinct magnetic-field-induced signatures associated with the spin-reorientation transition. In the temperature range of 60--90~K, the isotherms of the differential magnetic susceptibility, $\chi'(H)=dM/dH$, measured at different temperatures intersect within a very narrow magnetic-field interval, forming isosbestic points. Using the isosbestic-invariance approach, we establish a universal scaling behavior of $\chi'(H)$ throughout the spin-reorientation regime. Field-dependent MFM imaging provides real-space evidence for the accompanying evolution of the magnetic texture: within the conical state, stripe domains gradually transform into isolated bubble-like domains that emerge near the isosbestic field [see Fig. 3 (d) and (e)]. Our findings provide new insight into spin-reorientation phenomena in kagome magnets and identify crossing points and universal behavior of DMS as generic fingerprints of the conical state. More broadly, these results suggest a  general route to stabilize nontrivial magnetic textures through anisotropy competition in centrosymmetric magnets.

\textbf{Acknowledgments}: We acknowledge D. Vollhardt, L. Chioncel, F. Büttner and T. Karaman for fruitful  discussions and suggestions. This work was supported by the  Deutsche Forschungsgemeinschaft (DFG) through TRR (Project No. 360 - 492547816) and project ANCD (code 011201, Moldova). D.M.E. acknowledges and thanks funding from the DFG (grant
No. EV 305/1-1) and UK Research and Innovation (UKRI)
through Quantum Technology Fellowship [UKRI1221].

\appendix
\section{Universality of differential magnetic susceptibility for sample S2 }

Figure A1 shows the experimental data and their analysis for sample S2.

\renewcommand{\thefigure}{A\arabic{figure}}
\setcounter{figure}{0}
\begin{figure}[htp]
\centering
     \includegraphics[scale=0.08]{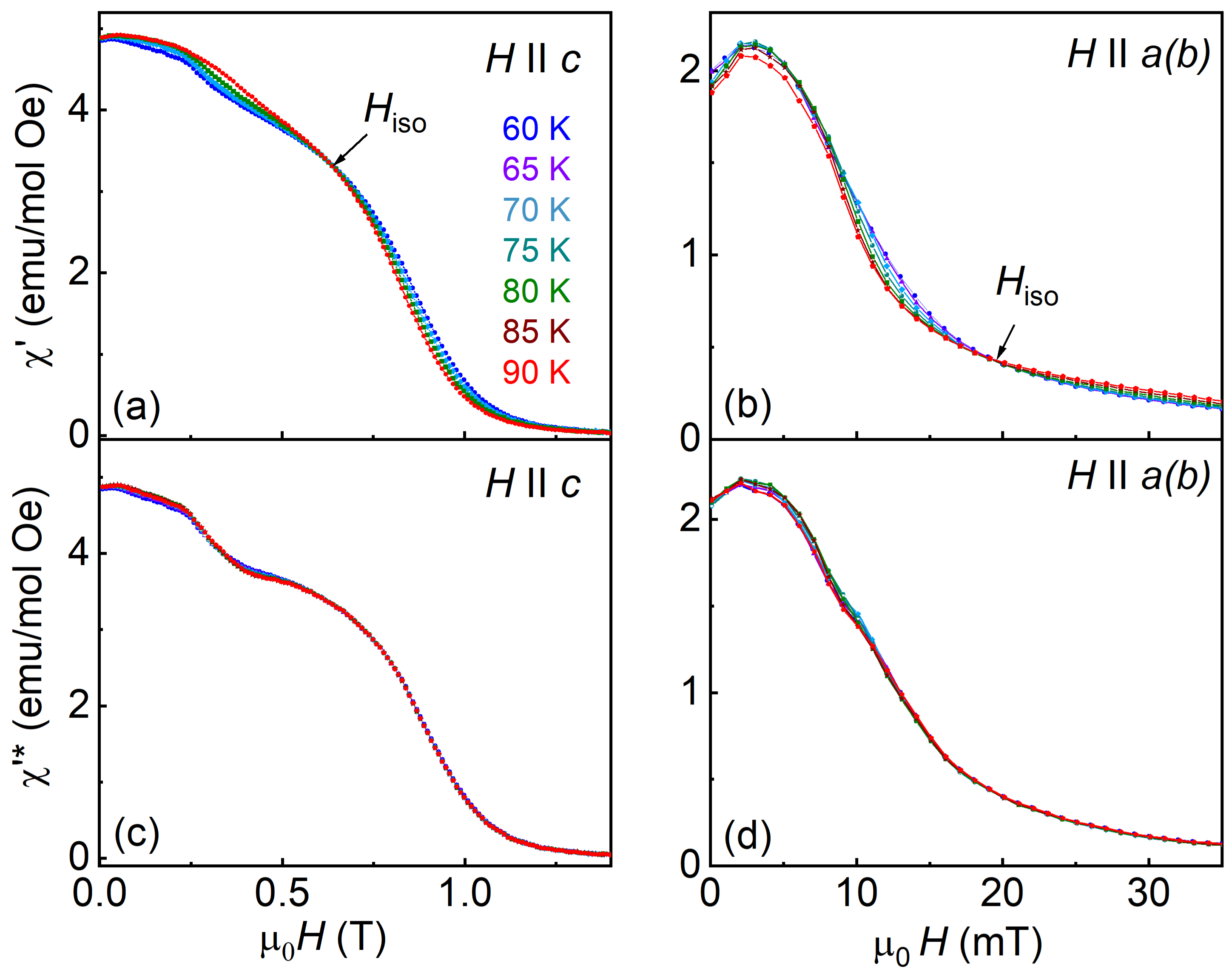}
    \caption{Differential magnetic susceptibility $\chi'$ as a function of magnetic field $H$ for a Fe$_3$Sn$_2$ single crystal (sample S2), measured in the temperature range 60--90~K with the magnetic field applied along the $c$ axis (a) and the $a$ axis (b). Panels (c) and (d) show the scaled quantity $\chi_i^{*}$, demonstrating the universal collapse of the corresponding isotherms onto a temperature-independent curve. }
  \label{figA1}
\end{figure}

\renewcommand{\thefigure}{A\arabic{figure}}
\setcounter{figure}{1}
\begin{figure}[htp]
\centering
     \includegraphics[scale=0.1]{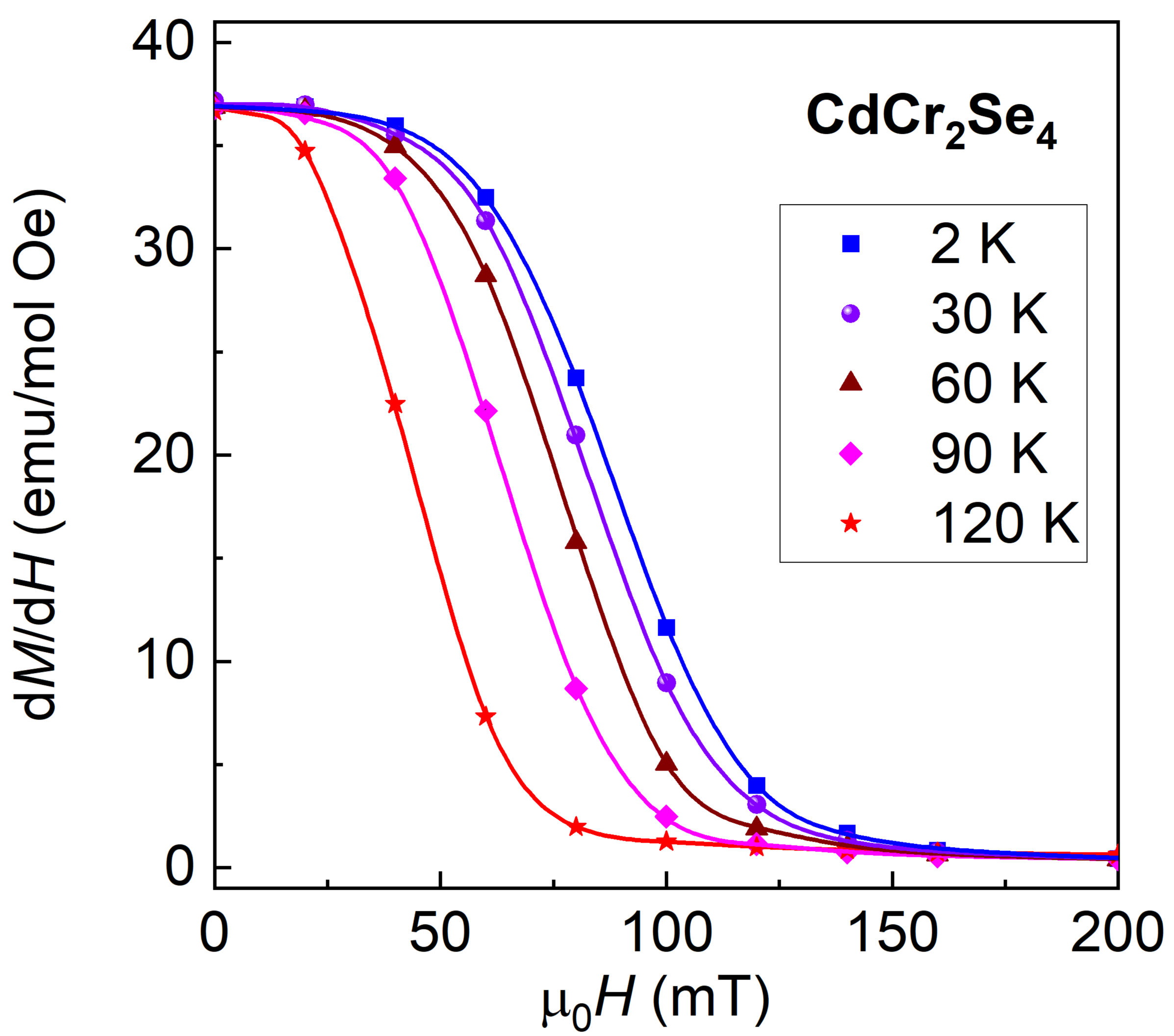}
    \caption{Isotherms of the differential magnetic susceptibility $\chi'$ as a function of magnetic field $H$ for a ferromagnetic CdCr$_2$Se$_4$ single crystal.}
  \label{figA2}
\end{figure}

\section{Differential magnetic susceptibility for isotropic and anisotropic ferromagnets }

Figure~A2 shows the differential magnetic susceptibility isotherms of the isotropic cubic spinel single crystal CdCr$_2$S$_4$.
Figure~A3 shows the DMS of the anisotropic easy-plane kagome ferromagnet Fe$_3$Sn, which does not exhibit a spin-reorientation transition.

\renewcommand{\thefigure}{A\arabic{figure}}
\setcounter{figure}{2}
\begin{figure}[htp]
\centering
     \includegraphics[scale=0.08]{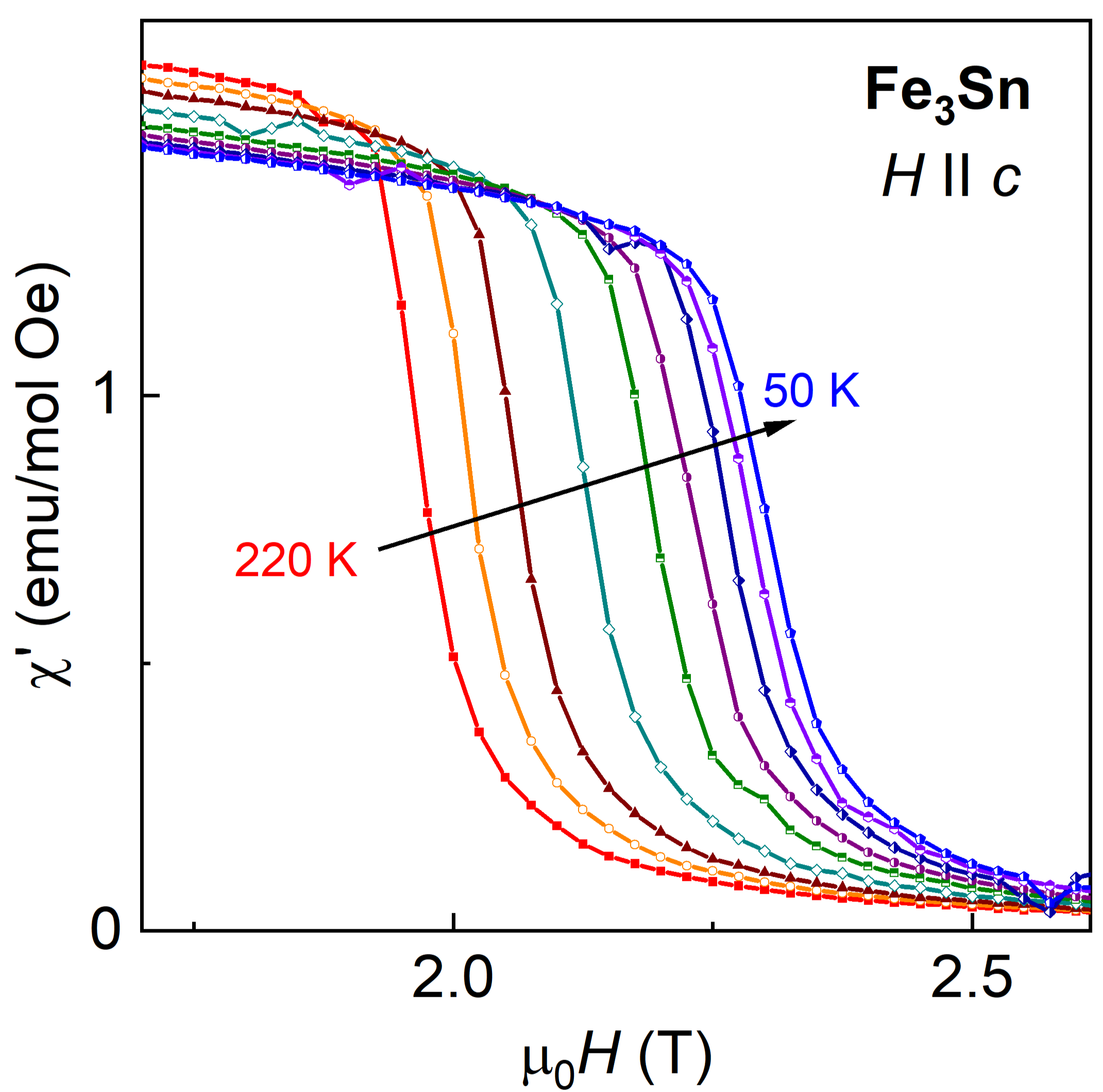}
    \caption{Isotherms of the differential magnetic susceptibility $\chi'$ as a function of magnetic field $H$ for a ferromagnetic Fe$_3$Sn single crystal. }
  \label{figA2}
\end{figure}

\bibliography{Ref}

\end{document}